\documentclass[epj]{svjour}
\usepackage{graphicx}
\usepackage{amsmath}
\usepackage{color}
\begin{document}
\title{Optical Hall conductivity of systems with
 gapped spectral nodes}

\author{Antonio Hill \inst{1}, Andreas Sinner\inst{1} \and Klaus Ziegler\inst{1} 
}                     
%
%
\institute{Institut f\"ur Physik, Universit\"at Augsburg,\\ Universit\"atsstra\ss{}e 1, D--86159 Augsburg, Germany}
\date{Received: date / Revised version: date}
%
\abstract{We calculate the optical Hall conductivity within the Kubo formalism for systems with gapped spectral nodes, where the latter have a power-law dispersion with exponent $n$. 
The optical conductivity is proportional to $n$ and there is 
a characteristic logarithmic singularity as the frequency approaches the gap energy. 
The optical Hall conductivity is almost unaffected by thermal fluctuations and disorder for $n=1$, whereas
disorder has a stronger effect on transport properties if $n=2$.
} 
\maketitle
\section{Introduction}
Transport properties of systems with two bands and spectral nodes are of great interest, as recent
studies of graphene 
have indicated. A prototype of this material class is
monolayer graphene (MLG) which is a monoatomic sheet of carbon atoms arranged in a honeycomb lattice with unique transport 
properties. This is a consequence of the  two--dimensional nature of the material and due to the band structure
which consists of two separate bands touching one another 
at isolated nodes. In the vicinity of these nodes quasiparticles exhibit a linear spectrum. 
The main difference between MLG and bilayer graphene (BLG) is that the low--energy excitations of the latter have
a quadratic spectrum~\cite{novoselov-blg,ohta}. For the longitudinal conductivity this difference causes a factor 
of 2 for the DC conductivity~\cite{cserti07,nicol} and also for the optical conductivity~\cite{abergel}. Additionally, we found a logarithmic singularity in the optical Hall conductivity for MLG in an earlier work~\cite{hill11}.
This leads to the question how a change of the low--energy spectrum around the nodes
affects the quantum Hall properties and the singularity, e.g., for a few layers of graphene.
For this purpose we generalize our model and assume that the spectrum in a small vicinity of the node is a power law with
integer $n$ ($n\ge 1$). 
We will call this spectral structure a node of order $n$.

An intriguing phenomenon in graphene is the quantum Hall effect (QHE) 
which was already observed in the first experiments on graphene~\cite{novoselov-mlg}. 
It exhibits a rather unexpected anomalous behavior. In contrast to the QHE in a
two--dimensional electron gas of a semiconductor, the Hall plateaux appear antisymmetrically around zero carrier
density~\cite{novoselov-blg}.  {The existence of these plateaux is commonly explained by a nonzero
Berry phase~\cite{novoselov-blg,novoselov-mlg,zhang-mlg,nagaosa10}.} Additionally, the magnitude of the Hall conductivity of
the first plateau is for BLG twice the corresponding value of MLG~\cite{novoselov-blg}. 
 {Similar effects appear in the case of rhombohedral (ABC) stacked trilayer graphene 
where the first plateau is found to be three times higher than in MLG~\cite{Zhang-tlg}.}
It is widely accepted that the QHE occurs in semimetals as a consequence of a gap opening and broken 
time reversal symmetry. Usually, this is achieved by applying a magnetic field perpendicular 
to the 2D plane, whereas the gap opening alone can be obtained by hydrogenation of MLG~\cite{elias09} 
or by a double gate in the case of BLG~\cite{ohta}.  
 {The optical Hall
conductivity was calculated for graphene in a homogeneous magnetic field~\cite{morimoto,gusynin2006,gusynin}.
}

 {
An alternative way for observing the QHE was suggested by Haldane, using a periodic magnetic flux on the
honeycomb lattice~\cite{haldane}. Such a periodic flux can be generated by a spin texture, realized by
spin doping of graphene \cite{hill11}. The main effect of the periodic flux is that the mass terms in the 
two valleys can be tuned independently. Then the QHE appears when the mass signs in the two valleys are different
because the Hall conductivity reads~\cite{hill11} 
\begin{equation}
\sigma_{xy}=\left[{\rm sgn}(m)-{\rm sgn}(m')\right]\frac{e^2}{2h}
\ .
\end{equation}
A similar case is a 3D topological insulator whose surface is covered with magnetically ordered spins
\cite{nomura}. The spins break the time-reversal invariance and open a gap in the surface Dirac cones.
}
\section{Model}
In the following we calculate the Hall conductivity for low--energy quasiparticles 
in the vicinity of a node with a uniform gap and a power--law spectrum with integer exponent.
Then the low--energy Hamiltonian, describing electrons in systems with spectral nodes with a uniform gap
$\Delta=2m$, reads in Fourier representation as 
\begin{equation}
H= \gamma \left(\begin{array}{cc}
     \gamma m & (k_x - ik_y)^n \\
    (k_x + ik_y)^n & - \gamma m
   \end{array} \right) \, ,
\label{hamiltonian}
\end{equation}
where $n=1$ should be associated with MLG and $n=2$ with BLG. For general order $n$ the eigenvalues of the
Hamiltonian are 
\begin{equation}
 E_l = (-1)^l E, \qquad  E= \gamma \sqrt{ (\gamma m)^2 + k^{2n}} \, ,
\end{equation}
where $l=0$ ($l=1$) refer to the upper (lower) band, respectively.
The band parameter $\gamma$ is for example in MLG $\gamma=v_F$~\cite{castro-neto} and in BLG $\gamma=v_F^2/\gamma_1$~\cite{mccann}, 
where $v_F$ is the Fermi velocity and $\gamma_1$ is the interlayer coupling constant.
To simplify the notation we drop indices for spin-- and valley degeneracy and put $\gamma$ equal to unity. 
The corresponding eigenvectors are
\begin{equation}
 \psi_{k}^{\pm}(\textbf{r}) =\sqrt{\frac{E\mp m}{2E}} \left( \begin{array}{c}
           \frac{(k_x - ik_y)^n}{ \pm E-m} \\ 1
          \end{array} \right) \exp(i \textbf{k}\cdot \textbf{r}) \, .
\end{equation}

{ If the stacking order of the layers is of type ABC, $n$ can be associated with the number of layers,
and the effective low-energy Hamiltonian is given by Eq. (\ref{hamiltonian}) \cite{min,zhang-abc-band}.}

\section{Optical Hall conductivity}
The Hall conductivity can be calculated as the off--diagonal element of the Kubo conductivity tensor:
\begin{align}
\nonumber 
\sigma_{\mu\nu} = \lim_{\alpha \rightarrow 0} \frac{i}{\hbar} \int \ \sum_{l,l'}\, 
\frac{\langle E_{l} | j_\mu | E_{l'} \rangle \langle E_{l'}| j_\nu |E_{l}\rangle}{E_{l} - E_{l'}} \\
\times  \frac{f(E_{l'}-E_F) - f(E_{l}-E_F)}{E_l - E_{l'} + \omega - i\alpha}  \frac{d^2 k}{(2\pi)^2} \, ,
\label{kubo-formel}
\end{align}
where $E_F$ represents the Fermi energy, $f(E)=1/(1+\exp(\beta E))$ the Fermi--Dirac distribution at the inverse temperature $\beta$ 
and $\omega$ the frequency of the external field. 
The current operator
\begin{equation}
j_\mu =ie [H,r_\mu] 
\end{equation}
with $r_\mu = i \partial/\partial k_\mu$ has vanishing diagonal elements. 
First we calculate current matrix elements defined in~(\ref{kubo-formel}). 
Due to rotational symmetry of the model, the use of polar coordinates is more convenient. 
Since the angular variable enters only the current matrix elements, the corresponding integration can be carried out separately. 
The intraband matrix elements
\begin{align}
\nonumber
\int_0^{2\pi} \langle\pm E| j_x| \pm E \rangle \langle \pm E| j_y| \pm E \rangle d\varphi &= \\
4e^2\,\int_0^{2\pi}  n^2 \, k^{4n-2} \, \frac{\cos(\varphi)\sin(\varphi)}{4E^2} \, d\varphi &= 0 \,
\end{align}
vanish after angular integration. 
The nonvanishing interband contribution of the matrix elements reads
\begin{align}
\nonumber \int_0^{2\pi} \langle\pm E| j_x| \mp E \rangle \langle \mp E| j_y| \pm E \rangle d\varphi &=\\
\nonumber \pm \int_0^{2\pi} i e^2 n^2  \frac{k^{2n-2}}{E^2}  \left \lbrace m E + i k^{2n} \cos(\varphi)\sin(\varphi) \right\rbrace  d\varphi &=\\
 \pm i 2\, \pi e^2 \, n^2 \, k^{2n-2} \, \frac{m}{E}& \, .
\label{angular-integration}
\end{align}
The current matrix elements are imaginary. In order to obtain the real part $\sigma_{\mu\nu}'$ of $\sigma_{\mu\nu}$ we have to evaluate a 
Cauchy principal value integral
\begin{align}
\nonumber
\sigma_{\mu\nu}' = \frac{i}{\hbar}  \int \ \sum_{l'\ne l }\,  2e^2\, i \pi \, n^2 \, k^{2n-2} \, \frac{m}{E_l} \, \frac{1}{E_{l} - E_{l'}} \\
\times\frac{f(E_{l'}-E_F) - f(E_{l}-E_F)}{E_l - E_{l'} + \omega } \ \frac{k \, dk}{(2\pi)^2} \, .
\end{align}
Substituting the $k$ integration by an integration over the energy $E=\sqrt{m^2+k^{2n}}$, the corresponding Jacobian becomes
\begin{equation}
  J =\left(\frac{\partial  E}{\partial k} \right)^{-1} =\frac{E}{n k^{2n-1}} \, .
\end{equation}
All powers of $k$ in the integrand cancel each other, such that the real part of the Hall conductivity reduces to
the simple expression
\begin{align}
  \sigma_{xy}' =&  \frac{n e^2 m}{ \pi\hbar} \int_{|m|}^{\infty} \,   \frac{f(-E-E_F)-f(E-E_F)}{4E^2-\omega^2} \,  dE \, .
\label{hall-integral}
\end{align}
The imaginary part is given by
\begin{align}
\nonumber
\sigma_{xy}'' =  \frac{n e^2 m}{ \pi\hbar} \int_{|m|}^{\infty} 
\frac{f(-E-E_F)-f(E-E_F)}{2E} \\ 
\label{imaginary-part}
\times\left[ \delta(2E + \omega)-\delta(2E- \omega)\right] dE .
\end{align}
In the limits $T\rightarrow0$ and $E_F \rightarrow0$ we obtain~\cite{hill11}
\begin{align}
\label{hall-cond-re}
\sigma_{xy}'=\frac{n}{2}\, \frac{e^2}{h}  \frac{m}{\omega} \ln\left| \frac{2m + \omega}{2m-\omega} \right|, \\
\label{hall-cond-im}
\sigma_{xy}''=-\, \frac{ne^2}{h}  \frac{m}{\omega} \, \theta(\omega - 2m)\, .
\end{align}
For $n=1$ and $n=2$ this result was also found independently by other authors~\cite{tse,nandkishore12,gorbar}
and reduces in the DC limit $\omega \rightarrow 0$ to
\begin{equation}
 \sigma_{xy}' = \mathrm{sgn}(m)\frac{n}{2}\, \frac{e^2}{h} \, , \ \ \ \sigma_{xy}'' = 0\, .
\label{hall-cond-DC}
\end{equation}
This DC result was also found for $n=1$ \cite{semenoff,sinitsyn}.
Hence, the DC Hall conductivity  is a nonzero constant, in units of $e^2/h$, which is proportional to the 
spectral exponent $n$.

The real part of the optical Hall conductivity increases with $\omega$ for $\omega<2m$ and 
decays like $\omega^{-1}$ for large $\omega$. 
 {
Remarkable is the singularity of $\sigma_{xy}'$ at $\omega=2m$ in~Eq. (\ref{hall-cond-re}). It appears when the 
external frequency $\omega$ is equal to the gap of the electronic system and was also discussed 
in the context of a band transition \cite{hill11,levitov11}. 
This point of the frequency spectrum separates two 
different regimes of the two-band model: For $\omega>2m$ the electronic system can absorb a photon from the 
external field and create a particle-hole pair, whereas this effect is forbidden for $\omega<2m$. 
The real part of the longitudinal conductivity undergoes a jump at $\omega=2m$ from $\sigma_{xx}'=0$
if $\omega<2m$ to $\sigma_{xx}'\approx e^2/h$ if $\omega>2m$ \cite{hill11}.
The divergent Hall conductivity can be understood as a combined effect of particle-hole creation and
an unhindered propagation of the particle-hole pairs. This propagation is given by the electronic two-particle 
Green's function $(H-\omega/2+i\epsilon)^{-1}(H+\omega/2-i\epsilon)^{-1}$ \cite{ziegler},
which has an exponential decay for $\omega<2m$, a power law decay at $\omega=2m$ and an oscillating
behavior for $\omega>2m$.   
The singularity of the optical Hall conductivity could be used to determine the gap experimentally
either by light \cite{nair} or synchrotron radiation \cite{li08}. 
}

\subsection{Finite temperatures}
\begin{figure}[ht]
\centering
\includegraphics[width=0.5\textwidth]{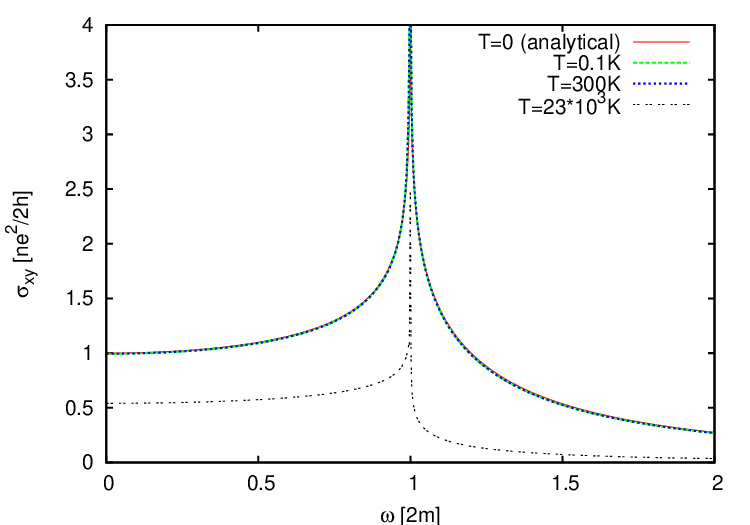}
\caption{Frequency and temperature dependence of the optical Hall conductivity.}
\label{temperatur-vergleich}
\end{figure}

\begin{figure}[ht]
\centering
\includegraphics[width=0.5\textwidth]{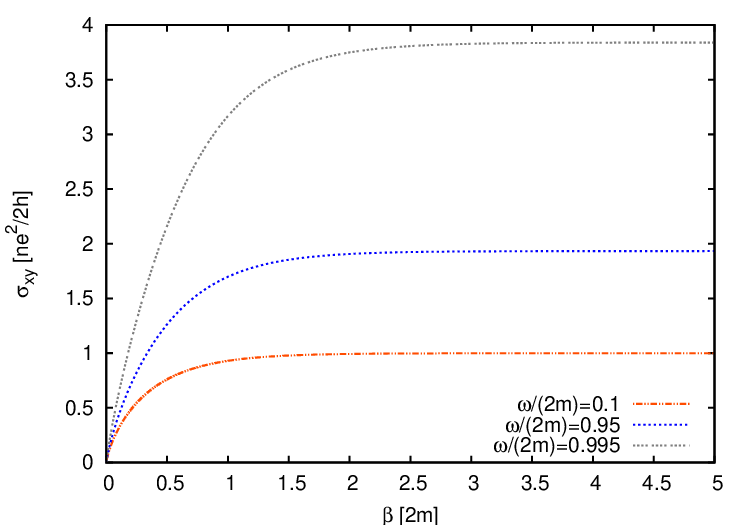}
\caption{Optical Hall conductivity near the singularity as a function of the inverse temperature $\beta$.}
\label{w-fest}
\end{figure}

In figure~\ref{temperatur-vergleich} we show a plot of expression~(\ref{hall-integral}) 
as a function of $\omega$ and different temperatures. The Hall conductivity scales with $n$.
This is in agreement with QHE experiments, where the plateau of BLG around zero carrier density 
is two times larger than in MLG~\cite{novoselov-blg,novoselov-mlg} and for trilayer graphene
three times larger than in MLG~\cite{Zhang-tlg}.
The temperature dependence is controlled by the energy scales of the system.
In graphene, a relevant energy scale is either the hopping parameter t $\approx 2.8eV$ or the gap with a similar energy, 
which corresponds to a temperature~$\approx 32.5 \times 10^3 K$. Therefore, expression~(\ref{hall-integral}) 
is insensitive over a wide range of temperatures $T\ll T_F$, as shown in figure~\ref{temperatur-vergleich}.
In particular, the gap singularity of the optical Hall conductivity survives. 
For very high temperatures, however, the Hall conductivity is reduced and goes eventually to zero. This is
shown for fixed frequency $\omega$ in figure  \ref{w-fest}.

\subsection{Disorder}
We have seen that thermal fluctuations have almost no effect on the singularity at $\omega=2m$. 
Since graphene
is also subject to disorder effects (ripples, impurities, etc.), we study their influence on the singularity in the
following.  {Complementary to \cite{nomura}, where the Hall conductivity of Dirac fermions in a random
vector potential is calculated numerically, disorder is introduced here via a scattering rate within the
self-consistent Born approximation.} 
This is a good approximation for one-particle properties such as the density of states, it fails for the DC ($\omega\to0$)
conductivity though due to singularities at $\omega=0$ \cite{ziegler07}. However, it is reliable again for the optical
conductivity if $\omega\gg\eta$ because the frequency plays the role of a cut-off for the singularities. The latter is also the
reason that the optical conductivity is not very sensitive to the type of disorder (e.g., scalar potential, vector potential or gap fluctuations)
and the details of the scattering type (e.g., intra- and inter-valley scattering).
Thus returning to the Kubo formula (\ref{kubo-formel}), we can rewrite the conductivity as (cf.~\cite{ziegler})
\begin{align}
\nonumber 
\displaystyle \sigma_{\mu\nu}=\lim_{\alpha\rightarrow0} \frac{i}{\hbar} &\int \int \frac{ \left\langle Tr\left[ j_\mu 
\delta(H_{dis}-\epsilon')j_\nu\delta(H_{dis}-\epsilon) \right] \right\rangle }{\epsilon - \epsilon' 
+ \omega -i\alpha} \\
 &\times \frac{f_\beta(\epsilon'-E_F)-f_\beta(\epsilon - E_F)}{\epsilon - \epsilon'} 
\ d\epsilon d\epsilon' \, ,
\label{kubo-disorder}
\end{align}
where $\langle...\rangle$ represents the disorder average. The latter can be approximated in the self-consistent Born approximation by replacing the Hamiltonian $H_{dis}$ by $\langle H_{dis}\rangle +i\eta$ \cite{abergel}. The average Hamiltonian $\langle H_{dis}\rangle$ is the same as the Hamiltonian in~(\ref{hamiltonian}) and $\eta$ is the scattering rate caused by the disorder. This implies that we have to replace the Dirac delta functions in~(\ref{kubo-disorder}) as
\begin{align}
\nonumber \delta (H_{dis}-\epsilon) \rightarrow \delta_\eta(H-\epsilon )=\\ 
\frac{i}{2\pi} \left[ (H-\epsilon + i\eta)^{-1} - (H-\epsilon - i\eta)^{-1} \right].
\end{align}
In MLG the scattering rate is $\eta \propto \exp(-\pi/g)$ \cite{ziegler-diffusion}, where $g$ is the variance of the
random gap. In BLG, on the other hand, $\eta$ is proportional to the variance $g$, which implies that the influence 
of disorder on BLG is much stronger. This difference is a consequence of the finite (divergent) density of states at the band edges of MLG (BLG). 
Realistic fluctuations in graphene near the Dirac node are less than a tenth of the hopping rate $g\approx 0.1$, which can be obtained by
extrapolating the measured data obtained away from the Dirac node \cite{tan07}. 
This results in scattering rates $\eta \approx 2\times 10^{-14}$ (MLG) and $\eta \approx 0.1$ (BLG). 

\begin{figure}[ht]
\centering
\includegraphics[width=0.5\textwidth]{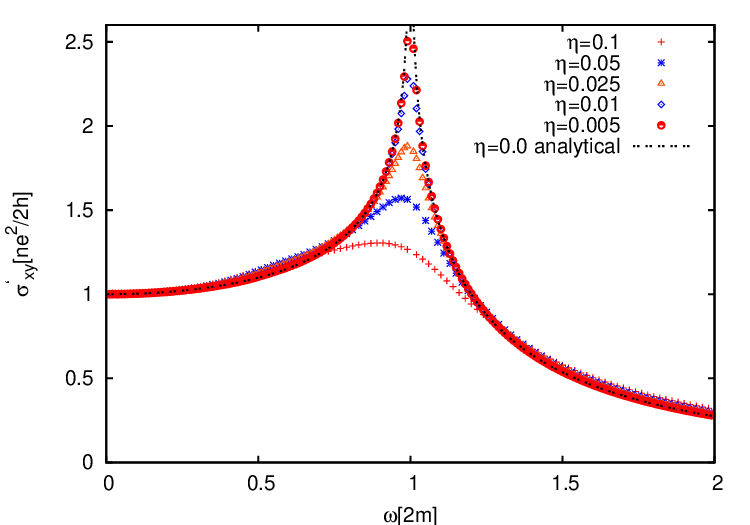}
\caption{Real part of the optical Hall conductivity for different values of the scattering rate $\eta$.}
\label{hall-cond-disorder}
\end{figure}

\begin{figure}[ht]
\centering
\includegraphics[width=0.5\textwidth]{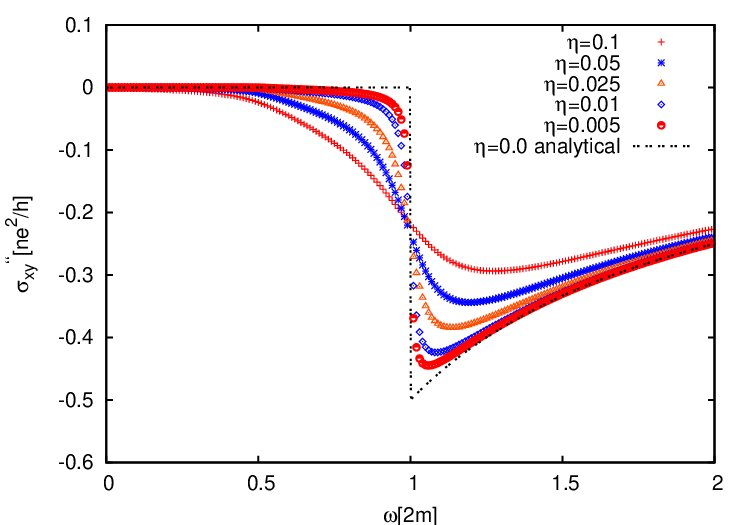}
\caption{Imaginary part of the optical Hall conductivity for different values of the scattering rate $\eta$.}
\label{hall-cond-dis-imag}
\end{figure}

Now the trace in Eq. (\ref{kubo-disorder}) can be expressed again in diagonal representation as
\begin{align}
\nonumber 
Tr\left[ j_\mu \delta_\eta(H-\epsilon')j_\nu\delta_\eta(H-\epsilon) \right] = \\
\nonumber \int \langle E_{l} | j_\mu | E_{l'} \rangle \langle E_{l'}| j_\nu |E_{l}\rangle \\ \delta_\eta(E_l-\epsilon')\delta_\eta(E_{l'}-\epsilon)  \frac{k\,dk}{(2\pi)^2} \, .
\end{align}
After angular integration (cf.~(\ref{angular-integration})) and transforming the $k$--integral to an energy 
integral the conductivity reads
\begin{align}
\nonumber \displaystyle
\sigma_{xy}' = \frac{e^2nm}{2\pi\hbar}\int \limits_{|m|}^{\infty} \int \limits_{-\infty}^{\infty}\int \limits_{-\infty}^{\infty}
\frac{f_\beta(\epsilon'-E_F)-f_\beta(\epsilon - E_F)}{(\epsilon - \epsilon')(\epsilon - \epsilon'+\omega)} \\
 \nonumber \times \left[\delta_\eta(E-\epsilon')\delta_\eta(-E-\epsilon) - \delta_\eta(-E-\epsilon')\delta_\eta(E-\epsilon)  \right] \\
\times d\epsilon  d\epsilon' dE.
\label{hall-cond-disorder-eq}
\end{align}
We evaluated expression~(\ref{hall-cond-disorder-eq}) for $T=0$ and $\mu=0$ numerically for several values of $\eta$. The results are depicted for the real and  imaginary part of the optical Hall conductivity in figures~\ref{hall-cond-disorder} and~\ref{hall-cond-dis-imag}. One can see that the Hall plateaux remains nearly constant for all $\eta$ under consideration, whereas the singularity is broadened.


\section{Conclusion}
In this work we have studied the optical Hall conductivity for systems with gapped nodes of order $n$. 
Our calculations indicate that the DC Hall conductivity for $n=1$ and for $n=2$ only depends on the sign
of the mass term and on the exponent of the low--energy spectrum. It reproduces the experimentally observed
factor of 2 for BLG. Interesting is that the optical Hall conductivity is quite insensitive to thermal 
fluctuations over a wide range of temperatures. The effect of the curvature of the low--energy spectrum is 
also surprisingly simple: The optical Hall conductivity is only multiplied by the factor $n$, as it was also found 
for $n=1,2$ in case of the longitudinal optical conductivity. It also reflects the $n$ dependence
of the visual transparency of multilayer graphene~\cite{nair}. Interestingly, there is a logarithmic singularity 
in the optical Hall conductivity when the frequency $\omega$ of the external AC field becomes equal to the gap 
of the electronic system. The appearance of the singularity in our calculations is related to the onset 
of particle--hole excitations for $\omega\ge 2m$. 
Although thermal fluctuations have no effect on this singularity, disorder may soften it in
the case of $n>1$, where the scattering rate $\eta$ can be large. In $n=1$ case, where the scattering rate 
is very small, the singularity of the optical Hall conductivity is almost unaffected. Consequently, the 
singularity could be used to determine the gap in MLG or topological insulators by measuring the optical Hall conductivity.

%

\end{document}